# External uniform electric field removing flexoelectric effect in epitaxial ferroelectric thin films

Hao Zhou[1], Jiawang Hong[2], Yihui Zhang[2], Faxin Li[1], Yongmao Pei[1(a)] and Daining Fang[1,2(b)]

[1] State Key Laboratory for Turbulence and Complex Systems, College of Engineering, Peking University – Beijing 100871, China
[2] Department of Engineering Mechanics, Tsinghua University - Beijing 100084, China

**Abstract** – Using the modified Landau–Ginsburg–Devonshire thermodynamic theory, it is found that the coupling between stress gradient and polarization, or flexoelectricity, has significant effect on ferroelectric properties of epitaxial thin films, such as polarization, free energy profile and hysteresis loop. However, this effect can be completely eliminated by applying an optimized external, *uniform* electric field. The role of such uniform electric field is shown to be the same as that of an ideal gradient electric field which can suppress the flexoelectricty effect completely based on the present theory. Since the uniform electric field is more convenient to apply and control than gradient electric field, it can be potentially used to remove the flexoelectric effect induced by stress gradient in epitaxial thin films and enhance the ferroelectric properties.

**Introduction.** –It is well known that ferroelectric thin films can be employed as the main components of nonvolatile random access memories for their switchable remnant polarization states [1-3]. With the rapid progress of nanoscale synthesis technology, the characteristic dimension of the films has fallen into nanoscale. However, with the thickness decreasing, the anomaly of many functional properties of ferroelectrics arises, such as the collapsed magnitude of dielectric constant [4], the increased remnant polarization [5] and piezoelectric coefficients [6], the reduced elastic stiffness [6]. One of the possible origins of these size effects is the inhomogeneous stress along the thickness direction of the ultrathin film, due to lattice mismatch between the ferroelectrics and substrate [7-11]. Actually, such inhomogeneous stress can affect performances of ferroelectric ultrathin film through two different mechanisms. One is due to the effect of stress itself [9-11], and the other is the influence of stress gradient.

So far, the knowledge of stress gradient effects (e.g. flexoelectricity, or FxE) is less clearer than that of stress effects (e.g. piezoelectricity), because the FxE is not obvious in most bulk dielectrics. The FxE was theoretically described about 50 years ago [12] and then its effect was discovered four years later by Scott [13] and Bursian et al [14]. However, the FxE effect has been overlooked due to its relatively small effect compared with piezoelectricity. Recently, the FxE effect has received increasing attentions because ferroelectric structures may undergo large stress gradient at nanoscale, which may have significant effect on properties [15,16]. In the past decade,

FxE is extensively investigated by experiments [15-22], first-principles [23-26] and macroscopic theory [8,27-30]. Lee et al reported that FxE in ferroelectric epitaxial nanofilms can be extremely large and, furthermore, can be modulated, which provides a means of tuning the ferroelectric properties such as domain configurations and hysteresis curves [21]. Catalan et al reviewed the studies on domains nanoelectronics, in which FxE also plays an important role due to the large strain gradient at domain walls [31]. However, to the best of our knowledge, there is still few practical methods to remove the FxE induced by the stress gradient in ferroelectric films, because it is difficult to apply or control a gradient field (such as electric field). In this paper, by using the generalized Landau–Ginsburg–Devonshire theory, which has repeatedly been found to be useful down to the nanoscale [32,33], we found that an external uniform electric field can almost eliminate the FxE effect on polarization, free energy profile and hysteresis loop completely in epitaxial thin films. The role of such uniform electric field is shown to be nearly the same as that of an ideal gradient electric field, while the uniform electric field is much more convenient to apply in practical applications.

**Theory and modeling.** – A *c*-phased heteroepitaxial, single-domain perovskite thin film ($P=P_3$) with misfit stress relaxing along the thickness direction is considered. Based on the phenomenological theory, the generalized free energy density function of thin film is given by

$$G=G_0+G_1+G_2+G_3 , \qquad (1)$$

(a) E-mail: peiym@pku.edu.cn
(b) E-mail: fangdn@pku.edu.cn

where $G_0$ is the free energy density of the paraelectric phase, $G_1$ the energy density which is the sum of electrostatic energy, elastic energy, depolarization field energy and surface energy, $G_2$ the FxE energy density, and $G_3$ external electric static energy density. $G_1$, $G_2$ and $G_3$ are given by

$$G_1 = \frac{1}{h}\int_0^h \left\{ \frac{1}{2}\alpha_0(T-T_{0\infty})P^2 + \frac{1}{4}\beta P^4 + \frac{1}{6}\gamma P^6 \right.$$
$$\left. -\frac{1}{2}(s_{11}+s_{12})X^2 - 2QXP^2 + \frac{1}{2}K\left(\frac{dP}{dz}\right)^2 \right. \quad (2)$$
$$\left. -\frac{1}{2}E_d P \right\} dz + \frac{K}{2}\left(\frac{P_s^2}{\delta_s} + \frac{P_i^2}{\delta_i}\right)$$

$$G_2 = \frac{1}{h}\int_0^h \left(-\zeta \frac{dX}{dz}P\right)dz, \quad (3)$$

$$G_3 = \frac{1}{h}\int_0^h (-E_e P)dz, \quad (4)$$

where $h$ is the film's thickness; $\alpha_0$, $\beta$ and $\gamma$ are dielectric stiffness and higher-order dielectric stiffness coefficients; $T_{0\infty}$ is the Curie-Weiss temperature of the bulk material; $\zeta$ describes the FxE coupling while $Q$ describes the electrostrictive coupling; $s_{11}$ and $s_{12}$ are the elastic compliance; $P_s$ and $P_i$ are the polarization at the surface ($z=h$) and the interface ($z=0$) while $\delta_s$ and $\delta_i$ are the corresponding extrapolation length; $E_e$ is the external electric field while

$$E_d = -\frac{1}{\varepsilon}\left(P - \frac{1}{h}\int_0^h P dz\right) \quad (5)$$

is the internal depolarization field [9], with $\varepsilon$ the dielectric constant of the film. A widely used model for the stress distribution is adopted as many other works [9,15,34]

$$X = X_0 e^{-kz}, \quad (6)$$

which means that the residual stress $X$ is exponentially relaxing from the film/substrate interface. $X_0$ is the interface residual stress, and it can be determined by the lattice constants $a_f$ of the film and $a_s$ of the substrate [10], as follows:

$$X_0 = \frac{(a_s - a_f)/a_s}{(s_{11}+s_{12})}. \quad (7)$$

The stress profile function Eq. (6) takes into account the relaxation difference among films of different thicknesses through the decline parameter $k$, given by [34] (unit: nm$^{-1}$)

$$k(h) = k_0 - \xi h = 3.925\times 10^{-3} - 2.325\times 10^{-6} h, \quad (8)$$

which increases as the film thickness decreases, describing a larger stress gradient state in thiner films, while $k=0$ represents a uniform stress state.

According to Eqs. (1-4), if an external electric field $E_e$ satisfies $G_2+G_3=0$, then the total free energy $G=G_0+G_1$. This provides us a possible way to remove the FxE theoretically by incorporating an external electric field in the energy function. Then we can obtain the following relationship:

$$\frac{1}{h}\int_0^h \left(-\zeta \frac{dX}{dz}P - E_e P\right)dz = 0, \quad (9)$$

Thereby, by applying the following electric field,

$$E_e(z) = -\zeta \frac{dX(z)}{dz} = k\zeta X_0 e^{-kz}, \quad (10)$$

the FxE effect can be removed thoroughly in the film. This requires the external electric field to be proportional to the stress gradient in the thickness direction of the thin film.

To obtain the polarization in the film, we carry out the variation of Eq. (1) with respect to $P$, yielding the following Euler's equation:

$$K\frac{d^2 P}{dz^2} = \left[\alpha_0(T-T_{0\infty}) - 4QX_0 e^{-kz} + \frac{1}{\varepsilon}\right]P + \beta P^3$$
$$+\gamma P^5 + k\zeta X_0 e^{-kz} - \frac{1}{h\varepsilon}\int_0^h P dz - E_e \quad (11)$$

with the boundary conditions at the surface and interface:

$$\begin{cases} \left.\frac{dP}{dz}\right|_{z=h} = -\frac{P}{\delta_s} \\ \left.\frac{dP}{dz}\right|_{z=0} = \frac{P}{\delta_i} \end{cases}. \quad (12)$$

The polarization distribution $P(z)$ in the thickness direction of the film can be obtained according to Eqs. (11-12) by using the finite-difference method.

**Numerical results and discussion.** – An epitaxial BaTiO$_3$ thin film is taken as an example in this paper. Parameters of the BaTiO$_3$ thin film used in the simulations are presented as follows [9,10,35,36]:
$\alpha_0 = 6.6\times 10^5$ VmC$^{-1}$, $\beta = 14.4(T-448)\times 10^6$ Vm$^5$C$^{-3}$,
$\gamma = 39.6\times 10^9$ Vm$^9$C$^{-5}$, $s_{11}+s_{12} = 5.62\times 10^{-12}$ m$^2$N$^{-1}$,
$K = 0.9\times 10^{-9}$ Vm$^3$C$^{-1}$, $\zeta = 2.69\times 10^{-9}$ m$^3$C$^{-1}$,
$Q = -0.043$ m$^4$C$^{-2}$, $T_{0\infty} = 383$ K, $\delta_i = \delta_s = 1$ nm.

Here we introduce the relative polarization $p(z)=P(z)/P_\infty$, where $P(z)$ is the polarization varying along the thickness direction of the film, and $P_\infty=\pm 0.27$ Cm$^{-2}$ is the spontaneous polarization value of the bulk BaTiO$_3$ counterpart. The positive and negative polarization corresponds to the polarization orientating towards the surface and interface, respectively. We adopt the misfit stress of $X_0=0.3$ GPa and thickness of $h=20$ nm, in accordance with other experimental work [37] and theoretical work [9, 30]. Based on our theoretical analysis and calculations, the proposed method to eliminate flexoelectric effect is applicable for a wide range of film thickness (from the theoretical critical thickness 15nm for reversible polarization to 1000nm). Note that the polarization due to flexoelectric effect would be less than 2% of the spontaneous polarization as the film thickness increases beyond 1000nm, and thus can be ignored.

According to Eq. (10), the external electric field to eliminate the FxE of the films is

$$E_e(z) = k\zeta X_0 e^{-kz}$$
$$= 3.130 \times 10^6 e^{-3.879 \times 10^{-3} z}. \quad (13)$$

The effects of such non-uniform electric field are shown in Fig. 1, which indicates that the FxE enhances the polarization all over the film as the black line with filled squares shows in Fig.1a. However, after applying external electric field as expressed in Eq. (13), the polarization profile (green line with hollow triangles) coincides with the original one without FxE and external electric field (blue line with hollow circles), which means the exponentially decreasing electric field can completely dispel the FxE effect on the polarization. However, such non-uniform electric field that is proportional to the stress gradient inside the film is difficult to apply and control in practical applications. As such, we propose to apply a uniform electric field instead, which is more convenient to apply and control, and also have nearly the same effect as the exponentially varying field.

If an external electric field $E_e$ is a uniform field $E_e^u$, then from Eq. (9), we will have

$$E_e^u = \frac{1}{h}\int_0^h \left(-\zeta \frac{dX}{dz}\right)dz = \frac{\zeta X_0}{h}\left(1 - e^{-kh}\right)$$
$$= 3.0 \times 10^6 \text{ V/m} \quad (14)$$

The effect of uniform electric field on the polarization profile is also shown in Fig. 1a (red line with filled triangles). As can be seen, it also overlaps with the line without FxE and electric field. The differences between polarizations without and with external exponential electric field (exp. EF) or uniform electric field (uni. EF) are shown in Fig.1b. It can be seen that the exponentially distributed electric field can completely eliminate the FxE effect everywhere inside the film (green line with filled squares), and uniform electric field has some discrepancies, especially at locations near two surfaces. However, these discrepancies are quite small (three orders of magnitude smaller than the spontaneous polarizations). Therefore, the uniform electric field is a good approximation for exponentially decaying electric field, and thus it can be employed to remove the FxE effect on the polarization in practical applications. In the following analysis, we will focus on the uniform electric field.

The elimination of FxE by external electric field can be explained from the energetic point of view. By using the finite-difference method, the energy profiles are obtained and shown in Fig. 2. It can be seen that, when there is no FxE coupling and no external electric field, the free energy is a symmetrical double-well profile as indicated by the blue line with hollow circles. However, FxE breaks this symmetrical double-well potential and results in an asymmetrical profile with negative

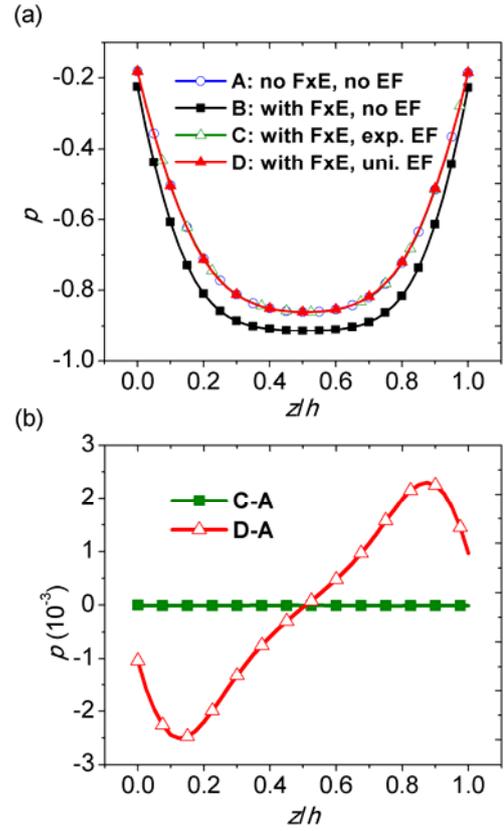

Fig. 1: (Colour on-line) (a) Polarization distribution in the thickness direction. (b) The differences between polarizations without and with external exponential electric field (green line with hollow triangles) or uniform electric field (red line with filled triangles). A, without FxE and external electric field; B, with FxE but without external electric field; C, with FxE and exponentially distributed electric field; D, with FxE and uniform electric field.

polarization (shown by the black line with hollow squares).This changes the double-well potential to single-well potential, and thus causes the increase of critical thickness of ferroelectric thin films [30]. Such effect resulted from FxE may hinder the application of ferroelectric devices at nanoscale. However, an external uniform electric field can remove this effect. Fig. 2 shows that the asymmetrical energy profile recovers to the symmetrical profile with the uniform positive electric field increasing gradually. When the electric field increases to $3\times10^6$ V/m (as shown in Eq.(14)), the effect of FxE is completely removed and the symmetrical double-well profile is recovered (see the overlap of red line with filled triangles and blue line with hollow circles). This means that the external uniform electric field can eliminate the energy profile distortion resulted from FxE. Similarly, the decrease of critical temperature [8,30] for reversible polarization induced by FxE can also be removed with the uniform electric field. If the electric field increases further, the asymmetry of energy profile emerges again with a positive polarization, as the dark cyan line with filled squares

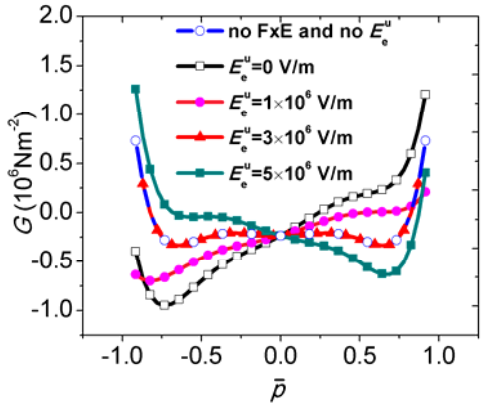

Fig. 2: (Colour on-line) Influence of various external uniform electric field on total free energy of epitaxial thin film at room temperature. Free energy of paraelectric state $G_0$ is set to zero.

shows in Fig.2.

*P-E* hysteresis loop is one of the most important properties of ferroelectric films, which has various applications in electric devices [2]. However, it was found that the relaxation of stress along the thickness direction inside the films can shift the hysteresis loop. As can be seen from Fig. 3 (black curve with filled squares), the FxE has a significant effect on the hysteresis loop, which shifts remarkably along the horizontal axis, and causes the increase of coercive field $E_{c2}$. More importantly, the change of coercive field due to FxE is greater ($E_{c2}/E_{c1} \approx 2.5$), as compared to the tiny increase of the polarization value due to the FxE effect (by ~1.1 times) (e.g. black line with filled squares and blue line with hollow circles in Fig.1a). Similar to the polarization and free energy profile, the shifted hysteresis loop due to FxE can also be moved back to the normal one (overlap of red curve with filled triangles and blue curve with hollow circles) by applying the same amount of uniform electric field $3 \times 10^6$ V/m. Thus, the shift of hysteresis loop induced by FxE in thin ferroelectric films can be eliminated by applying a proper uniform electric field and this could be used to improve the performance of ferroelectric devices.

Moreover, the large shift of hysteresis loop due to FxE may also be used to measure the FxE coupling coefficient $\zeta$. Figure 3 shows a proper uniform external electric field can move the hysteresis back to be normal loop and the required electric field is proportional to the FxE coupling coefficient and stress gradient, as Eq.(14) indicates. However, the stress distri-bution is complicated in the ferroelectric film-substrate system, so it is difficult to use this system to measure the FxE coupling coefficient. As we know, the stress in a pure bending beam is linearly distributed along thickness direction and its gradient is constant. This constant stress gradient distribution makes pure bending beam as a prototype system in the FxE coefficient measurement [18,20]. Using this pure bending beam system, the FxE coupling coefficient may also be measured by applying external electric field to remove the shift of hysteresis loop, as we discussed above. This method may avoid the dynamic

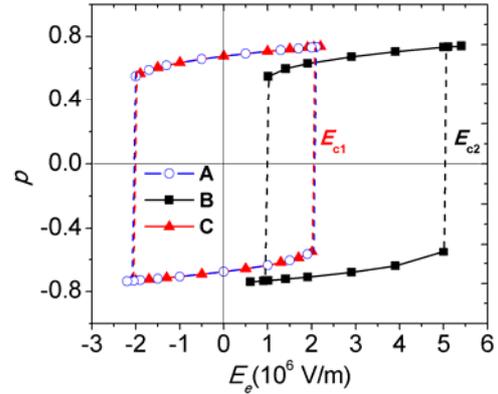

Fig. 3: (Colour on-line) Elimination of shift of hysteresis loop with an external uniform bias field. A, without FxE and external bias field; B, with FxE but without external bias field; C, with FxE and uniform bias field. $E_{c1}$ and $E_{c2}$ are the coercive electric field for A or C and B, respectively.

mechanical load and increase the accuracy of measurement.

**Conclusions.** – In summary, we found that the flexoelectric effect on polarization, free energy profile and hysteresis loop in epitaxial thin films can be removed by the use of a proper external electric field. An external uniform electric field plays almost the same role as that of an ideal gradient electric field which is proportional to stress gradient in thin films. Such uniform electric field is more convenient to apply and control in epitaxial thin films than gradient electric field, and thus can be used to eliminate flexoelectric effect in epitaxial films.

***

The authors are grateful for the support by National Natural Science Foundation of China under grants 11090330, 11090331 and 11072003. Support by the National Basic Research Program of China (G2010CB832701) is also acknowledged.